\documentclass{article}

\usepackage{arxiv}

\usepackage[utf8]{inputenc} 
\usepackage[T1]{fontenc}    
\usepackage{hyperref}       
\usepackage{url}            
\usepackage{booktabs}       
\usepackage{amsfonts}       
\usepackage{nicefrac}       
\usepackage{microtype}      
\usepackage{lipsum}		    
\usepackage{graphicx}
\usepackage{cleveref}   
\usepackage{doi}
\usepackage{siunitx}

\title{Certus Caliber Classification Gunshot Dataset (C3GD)}

\date{} 					

\author{ \href{https://orcid.org/0009-0008-4040-6474}{\includegraphics[scale=0.06]{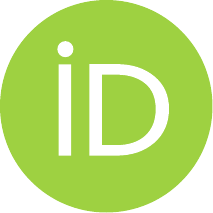}\hspace{1mm}Sinclair Gurny} \\
	Certus Innovations\\
	Albany, New York \\
	\texttt{sinclair.gurny@certusinnovations.com} \\
	\And
	\href{https://orcid.org/0009-0001-1946-2251}{\includegraphics[scale=0.06]{orcid.pdf}\hspace{1mm}Ryan Quinn} \\
	Certus Innovations\\
	Albany, New York\\
	\texttt{ryan.quinn@certusinnovations.com} \\
}

\renewcommand{\shorttitle}{Certus Caliber Classification Gunshot Dataset (C3GD)}

\hypersetup{
pdftitle={A template for the arxiv style},
pdfsubject={q-bio.NC, q-bio.QM},
pdfauthor={David S.~Hippocampus, Elias D.~Striatum},
pdfkeywords={Artificial intelligence (AI), caliber, deep learning, firearm, muzzle blast, signal processing},
}

\begin{document}
\maketitle

\begin{abstract}
    In this work, we introduce the Certus Caliber Classification Gunshot Dataset (C3GD), a publicly accessible dataset developed for the analysis of firearm muzzle blast sounds. The dataset aims to provide a wide variety of firearms, calibers, cartridges, microphones, and microphone locations with metadata detailed beyond what is currently available. It comprises more than 12,000 live-fire data points from 45 firearms across 18 calibers, including nine calibers not present in existing datasets. Live-fire data collection in semi-controlled environments is costly, but using internet-collected data increases the risk of low-quality data and label noise. This dataset focuses primarily on caliber classification, but can also facilitate gunshot detection, audio separation, and audio signal processing applications, providing a diverse and realistic reference. It aims to provide enough diversity to generalize to live-fire applications while also providing enough metadata for detailed academic analysis.
\end{abstract}


\section*{BACKGROUND}
Acoustic sensing offers distinct advantages over alternative sensing modalities. Unlike optical sensors, acoustic sensors do not require adequate lighting or line-of-sight, and audio surpasses many other sensor types in both range and directionality \cite{baiAcousticbasedSensingApplications2020}. Moreover, audio recording devices are widely available across existing infrastructure, including smartphones and security cameras, increasing accessibility. These properties make acoustic sensing particularly well-suited for gunshot detection and classification. Firearms produce loud, impulsive sounds with characteristic acoustic signatures \cite{naz_acoustic_2008} that propagate over long distances. The muzzle blast of a firearm travels considerably farther than the projectile itself, creating opportunities for real-time analysis in live-fire scenarios.

Automatic firearm classification from acoustic signatures has numerous applications in security operations, military contexts \cite{rakesh_sniper_2023}, forensic investigation \cite{maherAudioForensicInterpretation2025}, and public safety \cite{maoImprovingMassShooting2023}. The capability to distinguish not only the presence of gunfire but also characteristics such as caliber, firearm type, and firearm model can provide critical tactical intelligence for emergency response and threat assessment \cite{mares_gunshot_2022}. Consequently,  real-time acoustic classification models represent an opportunity to enhance safety and situational awareness in hostile environments.

Despite these compelling use cases, robust acoustic gunshot classification remains an open problem. The loud impulse caused by the muzzle blast can be mistaken for other loud sounds such as fireworks, construction work, and cars backfiring. Furthermore, its brief duration limits the temporal information available for analysis \cite{maher_modeling_2006}, presenting an even greater challenge on microphones that do not record at a high sample rate \cite{maher_acoustical_2007} \cite{lilien_development_2018}. To develop deep learning models that achieve robust generalization across diverse acoustic environments, the models must be trained on a variety of acoustic conditions including reverberation and echo, ambient background noise of varying signal-to-noise ratios, and signal distortion caused by microphone clipping \cite{naz_acoustic_2008}.

Commercial products for acoustic gunshot detection, such as ShotSpotter, demonstrate the difficulty of successfully detecting and classifying gunshots in uncontrolled conditions. ShotSpotter claims a \SI{>90}{\percent} accuracy rate, but independent investigations place the true positive rate at only \SI{13}{\percent}, with \SI{80}-\SI{92}{\percent} of alerts not confirmed to be shootings \cite{office_of_the_new_york_city_comptroller_nypds_2024}\cite{macarthur_justice_center_shotspotter_2021}. Published algorithms for gunshot detection report significant trade-offs between precision and recall \cite{bajzikIndependentChannelResidual2022} and classification algorithms report \SI{<90}{\percent} accuracy \cite{raponi_sound_2022}\cite{liFastIdentificationMethod2022}. Together, these results point to an open problem with widespread interest.

\subsection*{Firearm Nomenclature}
\begin{figure}[t]
  \centering
  \centerline{\includegraphics[width=\columnwidth]{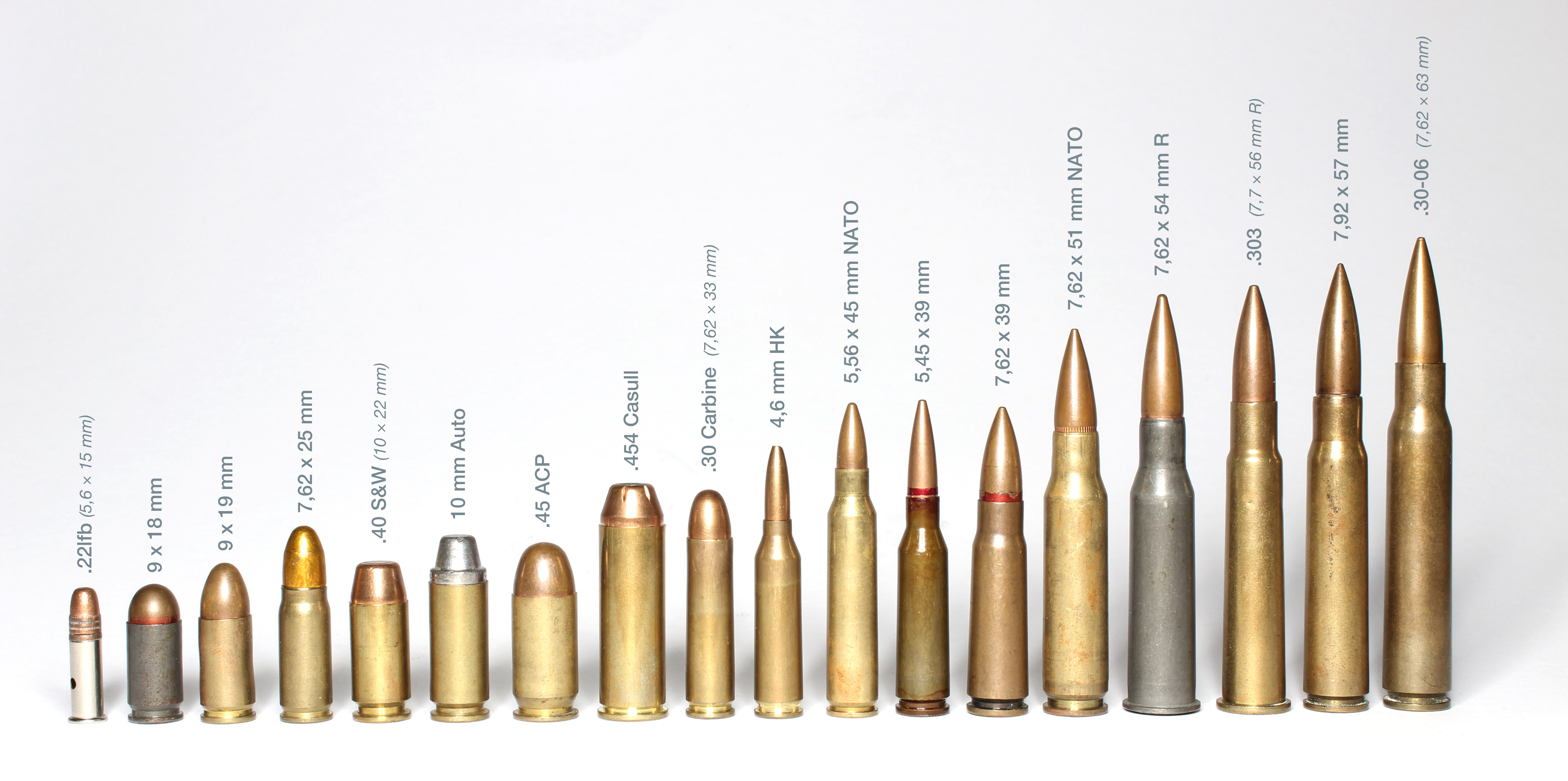}}
  \caption{Examples of standardized calibers. Image via \href{https://commons.wikimedia.org/wiki/File:Big_caliber_cartridge_comparison_withDATA.jpg}{Grasyl}}
  \label{fig:calibers}
\end{figure}

\textbf{Caliber} denotes the nominal bore or projectile diameter, typically expressed in inches or millimeters (e.g., .308 or 9mm). \textbf{Cartridge} refers to the fully assembled unit of ammunition composed of the casing, propellant, primer, and projectile, defined by precise dimensional and pressure standards. Firearms are chambered for specific cartridges, and cartridges sharing a nominal caliber are not necessarily interchangeable. Even very similar cartridges such as the .223 Remington and 5.56x45mm must be differentiated, as they conform to different pressure and manufacturing tolerances. Most cartridge standards are available with a variety of bullet weights, measured in grains, and constructions, such as hollow point or full metal jacket (FMJ). In this paper, caliber refers to specific ammunition standards as shown in \Cref{fig:calibers} (9x19mm), while cartridge defines the exact model and variation (e.g., 115gr FMJ).

\textbf{Platform} designates a specific configuration of firearm used for data collection. Many firearms are modular, so two firearms that share both make and model may differ significantly in barrel length and muzzle design.

\subsection*{Similar Work}
Despite myriad military, first responder, and public safety use cases, there is no publicly available gunshot audio dataset that provides consistent audio sample quality, a wide coverage of common firearm calibers, sufficient training data for deep learning models, and detailed annotations of firearm and cartridge metadata.

\begin{figure}[t]
  \centering
  \centerline{\includegraphics[width=\columnwidth]{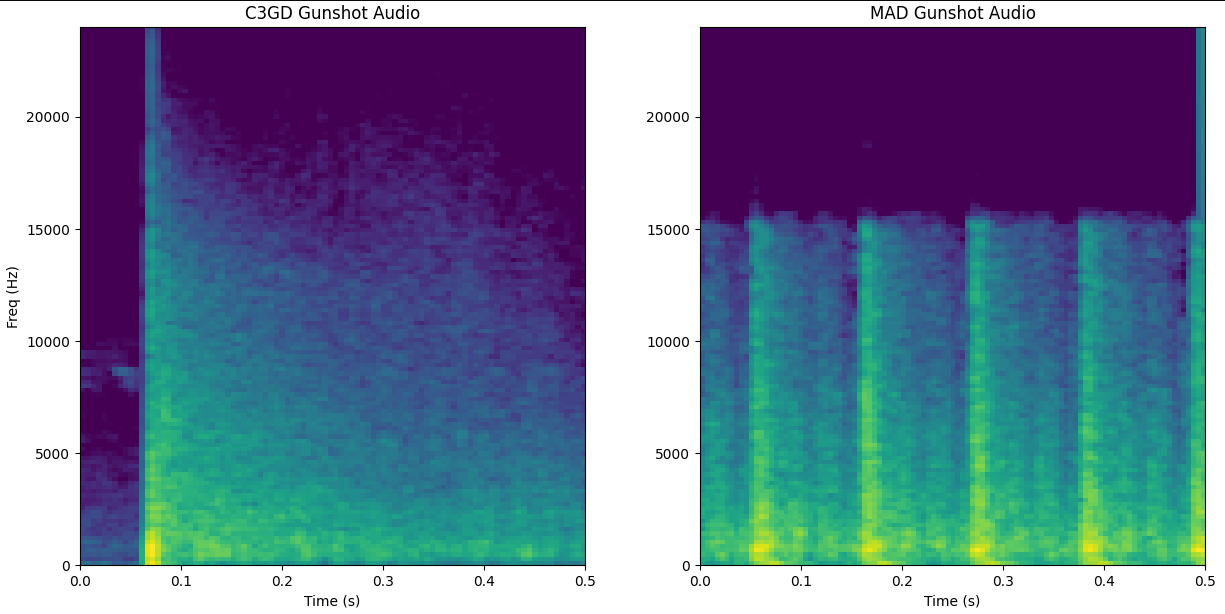}}
  \caption{Example of live-fire (left) vs. internet-collected (right) gunshot audio data}
  \label{fig:spectrum}
\end{figure}

Due to the cost and difficulty of collecting live gunshot data, many papers use data from YouTube \cite{raponi_sound_2022}\cite{tuncerAUTOMATEDGUNSHOTAUDIO2021}, Spotify \cite{shah_deciphering_2025}, and/or movie soundtracks \cite{bajzikIndependentChannelResidual2022}. Besides possible copyright issues with internet-scraped data, data quality is difficult to assure; audio data that has been compressed or down-sampled can never recover the lost high-frequency components. The authors have observed this loss of information in published datasets drawn from YouTube, notably the Military Audio Dataset (MAD) \cite{kimMilitaryAudioDataset2024}. Gunshot audio sampled at \SI{48}{\kilo\hertz} is expected to show content up to the Nyquist frequency, but many samples of gunshot audio from MAD lack content above \SI{16}{\kilo\hertz}, as seen in \Cref{fig:spectrum}. This loss of information is  seen in other datasets \cite{lilien_development_2018}\cite{zhaoHierarchicalTwoStageAudioResNet}.

A drop in both detection and classification accuracy when testing on internet-collected data has been noted by both \cite{lilien_development_2018} and \cite{bajzikIndependentChannelResidual2022}. Live-fire data collection avoids many of these limitations, but due to the expense and difficulty of collection, only two high-quality, general-purpose open datasets exist. 

\begin{itemize}
    \item Kabealo et al. (2023) \cite{kabealo_multi-firearm_2023} records ${\sim}3500$ samples across four calibers and platforms, including a rifle, pistol, revolver, and shotgun.
        \begin{itemize}
            \item \textbf{Strengths}: Diversity and locations of microphones, with 27 microphones at discrete locations around the shooter, including both smartphones and consumer-grade microphones. Provides at least one example of the four common form factors.
            \item \textbf{Limitations}: Limited coverage of firearms, particularly military calibers and hunting rifles. Only one firearm is used per caliber, and it does not provide metadata for cartridge types used, such as slugs or shot for the shotgun or sub-/supersonic ammunition for the 9x19mm pistol.
        \end{itemize}
    \item Lilien et al (2018) \cite{lilien_development_2018} recorded ${>}5100$ samples across at least 10 calibers and 18 platforms comprising multiple rifles, pistols, and revolvers.
    \begin{itemize}
        \item \textbf{Strengths}: Diversity of microphone locations, with four models (two phones and two dedicated audio recorders) placed at 20 different locations around the shooter. Caliber coverage is good for pistols and revolvers, and .22 LR is sampled from every available form factor (i.e., rifle, pistol, revolver, and carbine).
        \item \textbf{Limitations}: Lacks representation of both shotguns and hunting rifles. It does not distinguish between the usage of compatible calibers on the same platform, such as .223 Remington/5.56x45mm and .308 Winchester/7.62x51mm. Does not provide metadata for cartridge types used, such as sub-/supersonic ammunition for .22 LR or 9x19mm platforms.
    \end{itemize}
\end{itemize}

Other sources of gunshot audio are available for ML training, including UrbanSound8K \cite{salamonDatasetTaxonomyUrban2014} and the Free Firearm Sound Library \cite{jaszczak_free_2014}, both of which are used for gunshot detection in \cite{bajzikIndependentChannelResidual2022}. The former is not labeled by caliber/gun type and therefore cannot be used for classification, while the latter is not intended as a dataset and contains only 16--158 (median 30) samples per class.

C3GD roughly doubles the amount of freely-available labeled gunshot audio data and roughly triples the number of sampled firearm platforms. It introduces samples from nine new calibers, including three military calibers, from a total of 45 distinct platforms. Ten calibers have at least two platforms used, which allows investigation into differences introduced by form factor (e.g., pistol vs. carbine), barrel length, and manufacturer. For a detailed comparison of C3GD to existing datasets, see \Cref{tab:dataset_calibers}.

\begin{table}[htbp]
\centering
\small
\setlength{\tabcolsep}{4pt}
\begin{tabular}{l ccc ccc ccc}
 & \multicolumn{3}{c}{Kabealo et al.} & \multicolumn{3}{c}{Lilien et al.} & \multicolumn{3}{c}{C3GD} \\
\cmidrule(lr){2-4} \cmidrule(lr){5-7} \cmidrule(lr){8-10}
Caliber
 & \rotatebox{90}{Samples} & \rotatebox{90}{Platforms} & \rotatebox{90}{Cartridges}
 & \rotatebox{90}{Samples} & \rotatebox{90}{Platforms} & \rotatebox{90}{Cartridges}
 & \rotatebox{90}{Samples} & \rotatebox{90}{Platforms} & \rotatebox{90}{Cartridges} \\
\midrule
.22 LR             & --   & --  & -- & 1356 & 4  & --  & \textbf{1485}  & 6  & 6  \\
.223 Remington     & \textbf{1165} & 1   & -- & --   & -- & --  & 871   & 2  & 4  \\
.300 AAC Blackout       & --   & --  & -- & --   & -- & --  & \textbf{715}   & 2  & 3  \\
.357 MAG           & --   & --  & -- & \textbf{343}  & 1  & --  & --    & -- & -- \\
.380 ACP           & --   & --  & -- & 345  & 1  & --  & \textbf{775}   & 3  & 3  \\
.38 Special        & \textbf{702}  & 1   & -- & 336  & 1  & --  & 57    & 2  & 1  \\
.40 S\&W           & --   & --  & -- & \textbf{699}  & 2  & --  & 287   & 1  & 1  \\
.45 ACP            & --   & --  & -- & \textbf{1037} & 3  & --  & 513   & 2  & 3  \\
45-70              & --   & --  & -- & --   & -- & --  & \textbf{63}    & 1  & 1  \\
12 Ga              & \textbf{662}  & 1   & -- & --   & -- & --  & 477   & 4  & 3  \\
16 Ga              & --   & --  & -- & --   & -- & --  & \textbf{330}   & 1  & 1  \\
20 Ga              & --   & --  & -- & --   & -- & --  & \textbf{275}   & 1  & 4  \\
28 Nosler          & --   & --  & -- & --   & -- & --  & \textbf{25}    & 1  & 1  \\
5.56x45mm          & --   & --  & -- & --   & -- & --  & \textbf{1259}  & 4  & 4  \\
6.5 Creedmoor      & --   & --  & -- & --   & -- & --  & \textbf{913}   & 1  & 3  \\
7.62x39mm          & --   & --  & -- & \textbf{330}  & 1  & --  & 308   & 2  & 2  \\
7.62x51mm          & --   & --  & -- & --   & -- & --  & \textbf{280}   & 1  & 1  \\
7x57mm Mauser      & --   & --  & -- & --   & -- & --  & \textbf{140}   & 1  & 1  \\
9x19mm             & 1003 & 1   & -- & 702  & 2  & --  & \textbf{3339}  & 10 & 12 \\
\midrule
Total Samples      & 3532 & 4   &    & 5148 & 15 &     & 12112 & 45 & 54 \\
\end{tabular}
\caption{Sample counts by caliber across three datasets}
\label{tab:dataset_calibers}
\end{table}

C3GD is to the authors’ knowledge the only open gunshot audio dataset that records the cartridge fired to generate each sample. The authors expect this information to be relevant in both military force protection and first responder scenarios by, for example, differentiating military FMJ from law enforcement hollow-point rounds, or shotgun slugs vs birdshot or buckshot. Furthermore, the detail of the metadata allows for the possibility of third parties to extend the dataset using the same model and variations of firearm and cartridge, a level of reproducibility not possible from any other dataset.

Previous works \cite{lilien_development_2018}\cite{kabealo_multi-firearm_2023} have focused on first responder scenarios, where microphones are distributed around the shooter. This dataset is biased towards military and active shooter scenarios, where shots are generally traveling towards the microphones. This leads to many samples recording the ballistic shockwave of the bullet, which appear materially different on the consumer-grade hardware used in this work as compared to the high-end audio data collection system used in Nesar et al. (2024) \cite{nesar_machine_2024}.

\section*{COLLECTION METHODS AND DESIGN}

\subsection*{Class Selection and Coverage}

An authoritative list of the most common calibers or firearms was not found after an extensive search. As a best-effort proxy, statistical data from the United States Bureau of Alcohol, Tobacco, Firearms and Explosives (ATF) was combined with analytics from industry groups (e.g., Ronin's Grips Analytics) and publications by popular firearm websites such as The Gun Zone and Backfire.

While sampled calibers were limited by availability, the authors attempted to match class coverage with both the best available statistics and gaps in the existing open data ecosystem. Notably, the included calibers cover:

\begin{itemize}
    \item $>$85\% of manufactured pistols \cite{gordon_pistols_2024}
    \item $>$88\% of guns recovered by ATF law enforcement operations, as aggregated from their \href{https://www.atf.gov/resource-center/data-statistics/firearms-trace-data-reports?search_api_fulltext=&field_trace_data_location_parents=All&field_trace_data_location_children=All&field_report_end_date=+2024}{website}
    \item 9 of the 10 most popular calibers, per industry analytics \cite{roninsgrips_ranking_2025}
    \item The top 5 most popular calibers both for \href{https://thegunzone.com/what-are-the-top-selling-ammo-calibers/}{pistols} and \href{https://thegunzone.com/what-are-the-most-popular-handgun-calibers/}{overall}
    \item A majority of the most common \href{https://thegunzone.com/what-are-the-most-popular-rifle-calibers/}{rifle}, \href{https://thegunzone.com/what-is-the-most-popular-law-enforcement-handgun-caliber/}{law enforcement}, NATO, and \href{https://thegunzone.com/what-ammo-does-the-russian-military-use/}{Russian military} calibers
\end{itemize}

Hunting rifles are a coverage weakness of the dataset, with the popular .30-06 Springfield notably absent.

\subsection*{Data Collection}
The dataset was collected over four separate recording sessions conducted at distinct outdoor locations over a period of several months. Semi-controlled outdoor environments were selected to capture gunshot audio without excessive background noise while ensuring environmental variability. Samples with wind or animal noise were not excluded, but the dataset does not contain samples with notable vehicle noise or human speech. Images are included in the dataset that illustrate the microphone placement configurations for each collection event.

Each recording session employed a consistent protocol. Recording devices were positioned at intentional distance/azimuth from the shooter location, with primary placement in the downrange direction to capture acoustic propagation. The specifications and locations of each recording device are provided in the dataset metadata. Each firearm was fired using all the ammunition allotted to it, and the metadata was recorded manually upon conclusion of a section of testing, such as a particular platform or caliber. Each collection event employed two metadata recorders to limit errors and omissions.

The collection protocol encompassed 18 distinct calibers, with many calibers containing different ammunition variations, as well as multiple firearm makes and models to ensure diversity in barrel lengths, action types, and configurations. Each firearm was discharged multiple times to capture real-world variability in acoustic signatures while maintaining repeatability for reliable training data.

\subsection*{Data Processing}
Following each collection event, all recorded data was transferred from individual recording devices to a computer and imported into Audacity, an open-source audio editing software, for processing and quality assessment. The processing pipeline consisted of three primary stages: metadata annotation, multi-channel alignment, and temporal segmentation.

Metadata, such as gun model and caliber, was added to the data and the roughly-aligned data was exported as multi-track audio files for ease of processing, as a full day of recording can exceed \SI{40}{\giga\byte}. All audio, which was collected at or above \SI{48}{\kilo\hertz}, was resampled to a uniform \SI{48}{\kilo\hertz} sampling rate to ensure consistency across devices with varying native sampling rates.

Data was aligned across recording devices by identifying the muzzle blast onset in each recording manually. This step proved essential for comprehensive gunshot extraction, as lower-quality microphones or distant recording positions occasionally produced signals with insufficient amplitude for reliable automatic detection, as noted in \cite{kabealo_multi-firearm_2023}. Aligning all recordings to the highest-quality reference channel ensures consistent extraction across all devices while preserving the acoustic characteristics unique to each recording position and device type.

The aligned recordings were then processed using a custom Python script to automatically extract individual gunshot events. The extraction algorithm applies a first-order Butterworth high-pass filter with a \SI{7}{\kilo\hertz} \SI{3} dB point to the reference channel to suppress wind, human, and animal noise. It then searches for impulse peaks using a 100-sample rolling Z-score with a threshold of 3. Once a peak is found, it rolls back a selected period of time (about \SI{0.1}{\second}) to ensure the entire muzzle blast fits within the one-second window into which the data will be sliced. Each prospective sample is post-processed with four independent statistical checks that are strongly correlated with only gunshot audio (peak onset, SNR, spectral sharpness, spectral roughness). Any sample that failed at least one check was manually verified.

Once events were identified in the reference channel, the corresponding temporal segments were extracted from all aligned channels, ensuring that each gunshot was captured across all recording devices simultaneously. Using this gunshot extraction process, the audio was trimmed to non-overlapping clips that contain one or more distinct gunshots of a duration not to exceed one second. One second provides sufficient temporal context while maintaining computational efficiency for model training.

\section*{VALIDATION AND QUALITY}
All recordings underwent rigorous multi-stage quality control to ensure metadata accuracy, dataset reliability, and audio quality. The quality assurance process was integrated throughout data collection and processing phases. During collection events, field metadata was diligently recorded for each set of discharges, including firearm make, model, caliber, ammunition type, and any notable observations. Redundancy is built into the recording schema where the caliber and gun being fired is noted in the audio to ensure accuracy with written metadata. Recording sessions were conducted in semi-controlled conditions to minimize extraneous noise from vehicles, aircraft, and human speech. Firearm metadata was cross-referenced with manufacturer documentation and ballistic databases to minimize metadata errors. All collection events were conducted with a minimum of five microphones including at least two high quality reference microphones. All raw data was visually inspected using Audacity to ensure the presence of gunshots and absence of significant extraneous sounds. Furthermore, a portion of the final dataset was manually inspected to ensure quality.

\section*{RECORDS AND STORAGE}
The dataset is organized into several folders: data, metadata, and scripts. The data folder consists of 12,112 audio files in .wav format which follow the naming format described below. The metadata folder contains .csv files with detailed information about calibers, platforms used, exact cartridges used, microphone locations, microphone details, weather data, and testing events. Diagrams showing the exact microphone placements at each testing event are also included. The scripts folder contains example Python scripts for data extraction, feature generation, and benchmark model training.

\begin{center}
    \textit{ClassId-EventId-Platform-Mic-FileId-ClipId.wav}
\end{center}

Class ID is a unique identifier for each caliber, as listed in "classes.json." Event ID denotes where the data point was collected: OH Farm, NY Gravel Pit, or NJ Farm (note that two collection events were performed at the NJ Farm location). Platform is which of the 45 firearms was used to fire the sampled shot. File ID is a unique identifier for the original file the audio is clipped from, and clip ID is the ordinal clip number from the original file.

For basic replication, analysis, and comparison to other datasets, the top-level "metadata.csv" file maps audio files to metadata records, including class name, event, platform, cartridge, and mic. The values in these columns are foreign keys to the corresponding detailed metadata collections. An example test/train/val split is provided, which maintains class balance across all splits.

\section*{INSIGHTS AND NOTES}
\subsection*{Potential Use Cases}
While this dataset is specifically focused on caliber-based classification, it can be utilized for several sub-problems, such as platform or cartridge classification. Furthermore, in combination with external data, it can be used for gunshot detection.

Researchers should note that the semi-controlled recording environments lack the significant reverberation and background noise present in some real-world scenarios, and models trained on only this dataset may not generalize well without domain adaptations, but data with heavy reverberation or background noise may not generalize well either. To bridge this gap, we recommend applying data augmentations during training to simulate diverse conditions. Torch-compatible Python libraries are available for this purpose, including \href{https://github.com/iver56/audiomentations}{Audiomentations} and \href{https://pypi.org/project/teamml-audio-augment/}{teamml-audio-augment}, which provide a pipeline to simulate diverse environments by adding background noise, room impulse response/reverberation, and time-frequency masking. Incorporating these augmentations can reduce the performance gap between controlled recordings and real-world deployment conditions.

\subsection*{Benchmark}
To provide a reproducible benchmark for the dataset, the authors provide a Python script that trains and evaluates a deep learning model on the dataset with parameters taken from the existing literature. The model architecture is ResNet-18, as used in Bajzik et al (2022) \cite{bajzikIndependentChannelResidual2022}. Features are extracted from raw audio as log-mel spectrograms, which is common in modern gunshot detection and classification works \cite{bajzikIndependentChannelResidual2022} \cite{raponi_sound_2022} \cite{liFastIdentificationMethod2022} \cite{li_gunshot_2022}, with the specific spectrogram parameters taken from \cite{li_gunshot_2022}. Testing less common feature extraction methods, such as MFCCs, linear-frequency spectrograms, or raw audio, is beyond the scope of this work.

To assess basic reusability and generalizability benchmarks, the model was trained with both basic time shifting and a more robust data augmentation pipeline including time shifting, Gaussian noise, and gain adjustment. Time shifting is critical for this dataset to prevent the model from overtraining on the location of the gunshot within the sample window. Gaussian noise and gain augmentations provide a baseline simulation of realistic conditions such as varied distances and/or background noise. 

\begin{figure}[t]
  \centering
  \centerline{\includegraphics[width=\columnwidth]{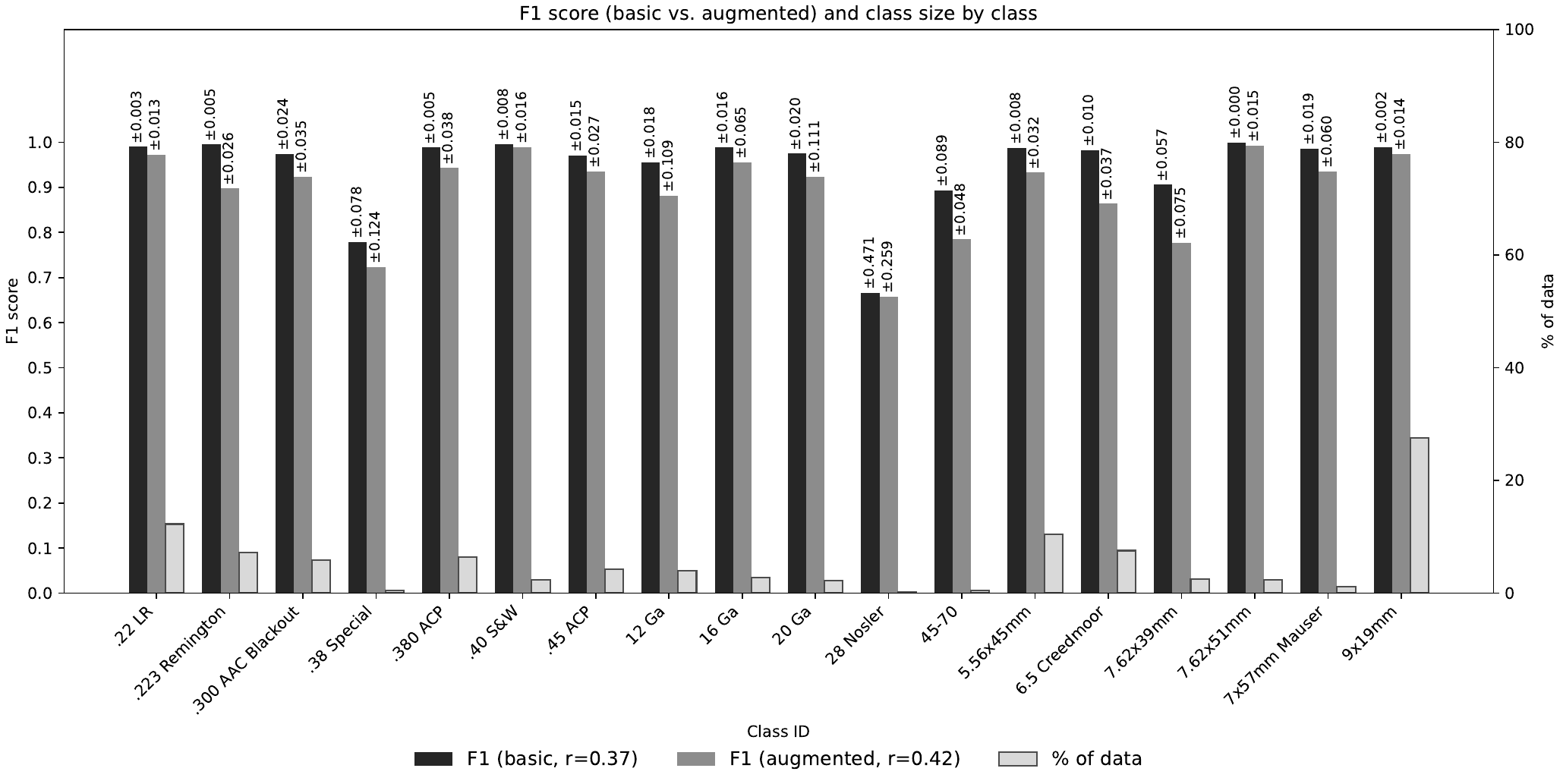}}
  \caption{Benchmark model results}
  \label{fig:results}
\end{figure}

Averaged across five runs, the basic model achieved a test accuracy of $98.35\% \pm 0.0045$ and the augmented model achieved $96.17\% \pm 0.028$. Detailed results across each class are displayed in \Cref{fig:results}. Notably, there is no significant correlation between class size and performance ($R^2 = 0.137$).

\subsection*{Limitations and Generalizability}
The dataset focuses on representing a realistic distribution of firearm classes sampled in multiple locations on heterogeneous consumer-grade hardware. However, the exact models and calibers are constrained by availability and may not fully represent all use cases. The lack of extensive hunting rifle data may make it poorly suited for anti-poaching use. Caution should be exercised when making real-world decisions based on machine learning models trained on only this dataset.

This dataset is presented with semi-clean audio, containing only ambient background noise. Since gunshot audio is very loud, it may be audible in situations with significant distortion and background noise. The authors recommend that users add relevant background noise during model training to augment the data. As discussed in the collection methods, the data collection locations consist of outdoor, field-like areas. Therefore, the dataset is unrepresentative of the echo and multipath present in urban or indoor environments. This limitation was an intentional choice based on the results of Routh and Maher \cite{routhRecordingAnechoicGunshot2016}, as muzzle blast analysis in urban and indoor environments is also a problem of acoustic propagation for which a dataset of this scale is insufficient to generalize across. Extensive impulse response datasets are available for data augmentation to address this limitation, for example, those listed on the \href{https://github.com/Graphi07/room-impulse-responses}{room-impulse-response GitHub page}

Despite these limitations, C3GD expands publicly available gunshot audio resources across multiple dimensions, using a wider range of microphones, platforms, and calibers than previous datasets.

\subsection*{Dataset Expansion}
Future work will focus on extending the dataset to include more calibers and firearms, particularly hunting rifles, and to include more detailed metadata, such as measured bullet speed and environmental data as it changes throughout the day. Furthermore, with higher-resolution environmental data, we aim to analyze the effects of air density and temperature. 

Supersonic ammunition produces a ballistic shockwave that originates from the bullet in motion \cite{maher_acoustical_2007}, which can be used for high-resolution shooter localization \cite{sallai_weapon_2011}. While C3GD records many samples of ballistic shockwaves, the time resolution is not sufficient for localization. No datasets currently contain labeled, annotated data on supersonic shockwaves. Future data collection should work to fill this gap for consumer-grade hardware.

\subsection*{Ethics Statement}
 This work did not involve human subjects, animal experiments, or data collected from social media platforms.

\section*{SOURCE CODE AND SCRIPTS} 
All training code, metadata, and documentation are available from \href{https://zenodo.org/records/22286299}{Zenodo} and are reproduced in \href{https://github.com/Stonewall-Defense/C3GD}{GitHub} for convenience. The scripts folder includes the dependencies, code, and configuration files necessary to reproduce the benchmark results. The GitHub README contains instructions for running the training script, which is also commented and annotated with Click commands.

The scripts/examples subdirectory contains example code showing how gunshot audio extraction and feature generation for models can be used in other contexts.

All internally developed tools for data extraction, analysis, and visualization are available on GitHub from \href{https://github.com/Stonewall-Defense}{Certus Innovations}. The authors used Audacity version 3.7.7 to preprocess the data.

The C3GD dataset and metadata are released under the \href{https://creativecommons.org/licenses/by/4.0/}{Creative Commons Attribution 4.0 International} license (CC-BY). Benchmark and example source code, as well as all dependencies developed by the authors, are released under the MIT license.

\section*{ACKNOWLEDGMENTS AND INTERESTS}
Sean Cyphert, Richard Newkirk, Peter Scialdo, Richard Brake, Kevan Dilworth, Chad Dennis, Nicholas Forezzi, and David Canestrare are acknowledged for data collection support. Special thanks to Travis Schuck and Andrew Finch for providing safe locations for data collection, and to Zev Nadler for volunteering as our range safety officer.

S.G. ran field data collection, helped design collection events, conducted all ML benchmark tests, and was the main author. R.Q. helped design collection events, post-processed and curated data, and wrote all supplemental code. All authors reviewed the manuscript.

This work was funded in part by Air Force Research Laboratory contract FA8750-24-C-B082.

The article authors have declared no conflicts of interest.

\section*{REFERENCES}

\bibliographystyle{ieeetr}
\bibliography{references}  

\end{document}